\documentclass[11pt,letterpaper]{article}
\pdfoutput=1

\usepackage{latexsym}
\usepackage{epsfig}
\usepackage{verbatim}
\usepackage{multirow}
\usepackage{fancybox}
\usepackage{shadow}
\usepackage{cite}
\usepackage{amscd, amsmath, amssymb, empheq}
\usepackage{bm, dsfont}
\usepackage{txfonts, yfonts, pmboxdraw}
\usepackage{graphicx}

\newtheorem{theorem}{\sc Theorem}[section]

\newcommand{\eps}{\varepsilon}

\newcommand{\proofend}{{\medskip\medskip}}
\newcommand{\proof}{{\noindent\em Proof. }}

\author{
  {\sc Bernard Chazelle}
\thanks{Department of Computer Science,
       Princeton University, 
{\tt chazelle}@{\tt cs.princeton.edu }}
}

\title{A Sharp Bound on the $s$-Energy and Its Applications
to Averaging Systems
\thanks{
The Research was sponsored by the Army
Research Office and the Defense
Advanced Research Projects Agency and was accomplished under Grant Number
W911NF-17-1-0078.
The views and conclusions contained in this document are those of the
authors and should not be
interpreted as representing the official policies, either expressed or
implied, of the Army Research Office,
the Defense Advanced Research Projects Agency, or the U.S. Government. The
U.S. Government is
authorized to reproduce and distribute reprints for Government purposes
notwithstanding any copyright
notation herein.
}}

\date{}

\begin{document} \maketitle

\vspace{.5cm}

\begin{abstract}
The {\em $s$-energy} is a generating function of wide applicability
in network-based dynamics.
We derive an (essentially) optimal bound of $(3/\rho s)^{n-1}$ on the $s$-energy of an $n$-agent
symmetric averaging system, for any positive real $s\leq 1$, where~$\rho$ is a lower bound on the
nonzero weights.  This is done by introducing the new dynamics of
{\em twist systems}. We show how to use
the new bound on the $s$-energy to tighten the convergence rates of
systems in opinion dynamics, flocking, and synchronization.
\end{abstract}

\vspace{1cm}

\section{Introduction}

Averaging dynamics over time-varying networks is a 
process commonly observed in 
many well-studied multiagent systems. It has been used
to model swarming, polarization, synchronization, gossip processes,
and consensus formation in distributed systems~\cite{bulloBk, FagnaniF, fortunato05}.
Because of a dearth of general convergence techniques,
results in the area often rely on network connectivity assumptions. 
The {\em $s$-energy} is a powerful analytical tool that 
allows us to overcome these restrictions~\cite{chazelle-total}.
It provides a global parametrized measure of the ``footprint" of the
system over an infinite horizon. 
This stands in sharp
contrast with the local arguments (spectral or Lyapunov-based) typically
used to prove fixed-point attraction. 

The main result of this paper is an optimal bound on the $s$-energy of
symmetric averaging systems.
The new bound is used to tighten the convergence rates
of various multiagent systems in opinion dynamics, flocking, and 
self-synchronization of coupled 
oscillators~\cite{bulloBk, chazelle-total, CuckerSmale1, hegselmanK, HendrickxB, jadbabaieLM03, jadbabaieMB, lorenz05, Moreau2005, strogatz00, tsitsiklis84, vicsekCBCS95}.

Before moving to the technical discussion,
we illustrate the role of the $s$-energy with
a toy system. Fix $\rho\in (0, 1/2]$
and place $n$ agents at $x_1,\ldots, x_n$ in $[0,1]$.
Given any $\eps>0$,
for any integer $t>0$, pick two agents $i,j$ 
such that $x_j-x_i\geq \eps$ (if any) and move 
them anywhere in the interval
$[x_i+ \delta, x_j-\delta]$, where $\delta= \rho(x_j-x_i)$. 
Repeat this process as long as possible.   Note the high nondeterminism of
the dynamics: not only can we choose the pair of agents
at each step, but we can move them anywhere we please within
the specified interval.  Despite this freedom, 
the process always terminates 
in $O\bigl( \frac{1}{\rho n}\log \frac{1}{\eps}\bigr)^{n-1}$ steps,
for any small enough $\eps>0$, and the bound is 
tight.\footnote{All logarithms are to the base~2.}
This result is a direct consequence of our new bound on the $s$-energy.
The proof relies on a reduction to {\em twist systems}, a new type
of multiagent dynamics that we define in the next section.

\paragraph{The $s$-energy.}

Let $(g_t)_{t=1}^{\infty}$ be an infinite sequence of graphs over
a fixed vertex set $\{1,\ldots, n\}$. Each $g_t$ is
embedded in $[0,1]$, meaning that its vertices (the ``agents") are 
represented by $n$ real numbers between 0 and 1.
Let $\mu_1,\ldots, \mu_k$ denote the lengths of the intervals
formed by the union of the embedded edges of $g_t$,
and put $\ell_t= \mu_1^s+\cdots+ \mu_k^s$, for real or complex $s$.\footnote{For example,
if $g_t$ consists of three edges embedded as $[0,0.2]$, $[0.1,0.3]$, $[0.7, 0.9]$,
and one self-loop at $0.5$, then the union of the edges forms the
three intervals $[0,0.3]$, $[0.5,0.5]$, $[0.7, 0.9]$
and $\ell_t= (0.3)^s+ (0.2)^s$.}
The {\em $s$-energy} ${\mathcal E}(s)$ of the system is defined as the infinite sum
$\sum_{t> 0} \ell_t$.
Because the $s$-energy follows an obvious scaling law, we note that
embedding the graphs in the unit interval is not restrictive.

\paragraph{Averaging systems.}

In a (symmetric) {\em averaging system},
$g_t$ is undirected and supplied with self-loops at the vertices.
To simplify the notation, we fix $t\in \mathbb{Z}^+$ 
and denote by $x_i$ and $y_i$ the positions of vertex $i$
at times $t$ and $t+1$, respectively.
Vertices are labeled so that $x_1\leq \cdots \leq x_n$.
For each $i\in \{1,\ldots, n\}$, write $r(i) = \max\{j  \,| \, (i,j)\in g_t \}$
and $l(i) = \min\{j \,| \, (i,j)\in g_t \}$.\footnote{Because $g_t$ is
undirected and has self-loops, $l(i)\leq i\leq r(i)$, 
$(l \circ r)(i)\leq i\leq (r \circ l)(i)$.
The notation $l, r$ should not obscure the fact that both functions can be chosen
differently for each graph $g_t$ and its embedding $(x_i)_{i=1}^n$.}
Fix $\rho\in  (0,1/2]$.
The move of vertex $i$ from $x_i$ to $y_i$ is subject to
\begin{equation}\label{iter}
x_{l(i)}+ \delta_i \leq y_i \leq  x_{r(i)} - \delta_i,
\end{equation}
where $\delta_i= \rho( x_{r(i)}-x_{l(i)} )$.
In other words, vertex $i$ can move anywhere within the interval covered by
its incident edges, but not too close to the endpoints. If $\rho=0$,
convergence is clearly impossible to ensure since $i$ can easily oscillate periodically
between two fixed vertices. We emphasize the high nondeterminism
of the process: $g_t$ is arbitrary and so is the motion of $i$ within its
allotted interval.

\paragraph{The results.}

Although the $0$-energy is typically unbounded,
it may come as a surprise that ${\mathcal E} (s)$ 
is always finite for any $s>0$~\cite{chazelle-total}.
In particular, the case $s=1$ shows that it takes only a finite amount of ink to
draw the infinite sequence of graphs $g_t$. We state the main result of this 
article,\footnote{We actually prove the slightly
stronger bound of $2(2/\rho s)^{n-1}$ for $n>2$.}
and prove it in~\S\ref{BoundsEnergy}:

\begin{theorem}\label{s-energyBound}
$\!\!\! .\,\,$
The $s$-energy satisfies ${\mathcal E}(s)\leq  (3/\rho s)^{n-1}$,
for any $0<\rho\leq 1/2$ and $0<s\leq 1$.
\end{theorem}

We prove in~\S\ref{lbProofs}.A that the bound $O(1/\rho s)^{n-1}$ is optimal 
for $s = O\bigl( 1/\log \frac{1}{\rho} \bigr)$ and $\rho \leq 1/3$.
These are the conditions we encounter in practice, which is why  
we are able to provide tight bounds for all the applications discussed in this work.
For $s=1$, a quasi-optimal lower bound
of $\Omega(1/\rho)^{\lfloor n/2 \rfloor}$ is already known~\cite{chazelle-total}.
Theorem~\ref{s-energyBound} lowers the previous upper bound of 
$(1/s)^{n-1}(1/\rho)^{n^2+ O(1)}$~\cite{chazelle-total}.

The $s$-energy helps us bound the convergence rates of averaging 
network systems in full generality. To our knowledge, 
no other current technique can prove these results. 
The power of the $s$-energy is that it makes no connectivity requirements about
the underlying dynamic networks. We use it typically to bound
the {\em communication count} $\mathcal{C}_\eps$,
which is defined as the maximum number of
steps~$t$ such that $g_t$ has at least one edge of length $\eps>0$ or higher.
From the inequality $\mathcal{C}_\eps\leq \eps^{-s} {\mathcal E}(s)$,
setting $s=1/\log\frac{1}{\eps}$
and $s=n/\log\frac{1}{\eps}$ in Theorem~\ref{s-energyBound} yields:

\begin{theorem}\label{CommCount}
$\!\!\! .\,\,$
The communication count satisfies
$\mathcal{C}_\eps = O \Bigl( \frac{1}{\rho} \log \frac{1}{\eps} \Bigr)^{n-1}$
for any $2^{-n}\leq \eps\leq 1/2$, and  
$\mathcal{C}_\eps = O \Bigl( \frac{1}{\rho n} \log \frac{1}{\eps} \Bigr)^{n-1}$
for $0<\eps<  2^{-n}$.
\end{theorem}

This lowers the previous upper bound of 
$(1/\rho)^{n^2+O(1)} (\log 1/\eps)^{n-1}$~\cite{chazelle-total}.
We prove in~\S\ref{lbProofs}.B that the new
bound is optimal for any positive $\eps\leq \rho^{2n}$ and $\rho\leq 1/3$.
We close this introduction with a few
remarks about the results and their context:   

\begin{enumerate}
\item
The results extend to a large family of asymmetric averaging systems.
Indeed, Theorems~\ref{s-energyBound} and~\ref{CommCount} hold for
any infinite sequence of cut-balanced digraphs $g_t$:  recall that a directed graph is said
to be {\em cut-balanced} if its weakly connected components
are also strongly connected. 

\item
The polylogarithmic factor 
$\bigl(\log 1/\eps \bigr)^{n-1}$ in the convergence rate of Theorem~\ref{CommCount}
is a distinctive feature of time-varying network-based dynamics.
Markov chains, for example, have convergence rates 
proportional to $\log 1/\eps$.

\item
Our definition of the $s$-energy differs slightly from the
original formulation~\cite{chazelle-total}, which introduced
the {\em total $s$-energy} as
$\sum_{t> 0} \sum_{(i,j)\in g_t} d_{ij}(t)^s$, where
$d_{ij}(t)$ is the distance between the vertices $i,j$ in the embedding of $g_t$.
Up to a correction factor of at most  $\binom{n}{2}$,
our bounds apply to the total $s$-energy as well.

\item
As noted in~\cite{chazJSP}, 
the $s$-energy can be interpreted
as a generalized Dirichlet series or,
alternatively, as a partition function 
with $s$ as the inverse temperature.
Both interpretations have their own benefits, such as highlighting
the lossless encoding properties of the $s$-energy 
or the usefulness of Legendre-transform arguments 
with the relevant thermodynamical quantities.
\end{enumerate}

\section{Twist Systems}

We reduce averaging systems to a simpler kind of dynamics
where agents keep the same ordering at all time.
In a {\em twist system}, $n$ points move within $[0,1]$ at discrete
time steps. As before, we fix $t\in \mathbb{Z}^+$ and 
describe the motion of each point $x_i$ at time $t$ to its next 
position $y_i$ at time $t+1$.  Unlike the averaging kind, 
twist systems preserve order; that is, assuming that
$x_1\leq \cdots \leq x_n$, then $y_1\leq \cdots \leq y_n$.
To describe the motion from $t$ to $t+1$, we 
choose two integers $1\leq u<v \leq n$
and, for any $i$ ($u\leq i\leq v$), we define
the {\em twist} of $x_i$ as
the interval within $[x_u,x_v]$ defined by
\begin{equation}\label{twistDef}
\tau_i = \bigl[x_u + \rho (x_{\min\{i+1, v\}}-x_u),  x_v- \rho (x_v- x_{\max\{i-1,u\}})\bigr].
\end{equation}
Fixing $\rho\in (0,1/2]$ ensures that
all the twists are well-defined.\footnote{Indeed, we can check that
$\tau_i= [a,b]$, where $a\leq b$.
The terminology refers to the ``twisting" of the interval $[x_{i-1}, x_{i+1}]$
around $x_i$ into the interval $\tau_i$ around $y_i$.}
The only constraints on the dynamics are: (i) $y_1\leq \cdots \leq y_n$;
and (ii) $y_i\in \tau_i$ for any $u\leq i\leq v$, and $y_i=x_i$ otherwise.

\vspace{-0.3cm}
\begin{figure}[htb]
\begin{center}
\hspace{0cm}
\includegraphics[width=7cm]{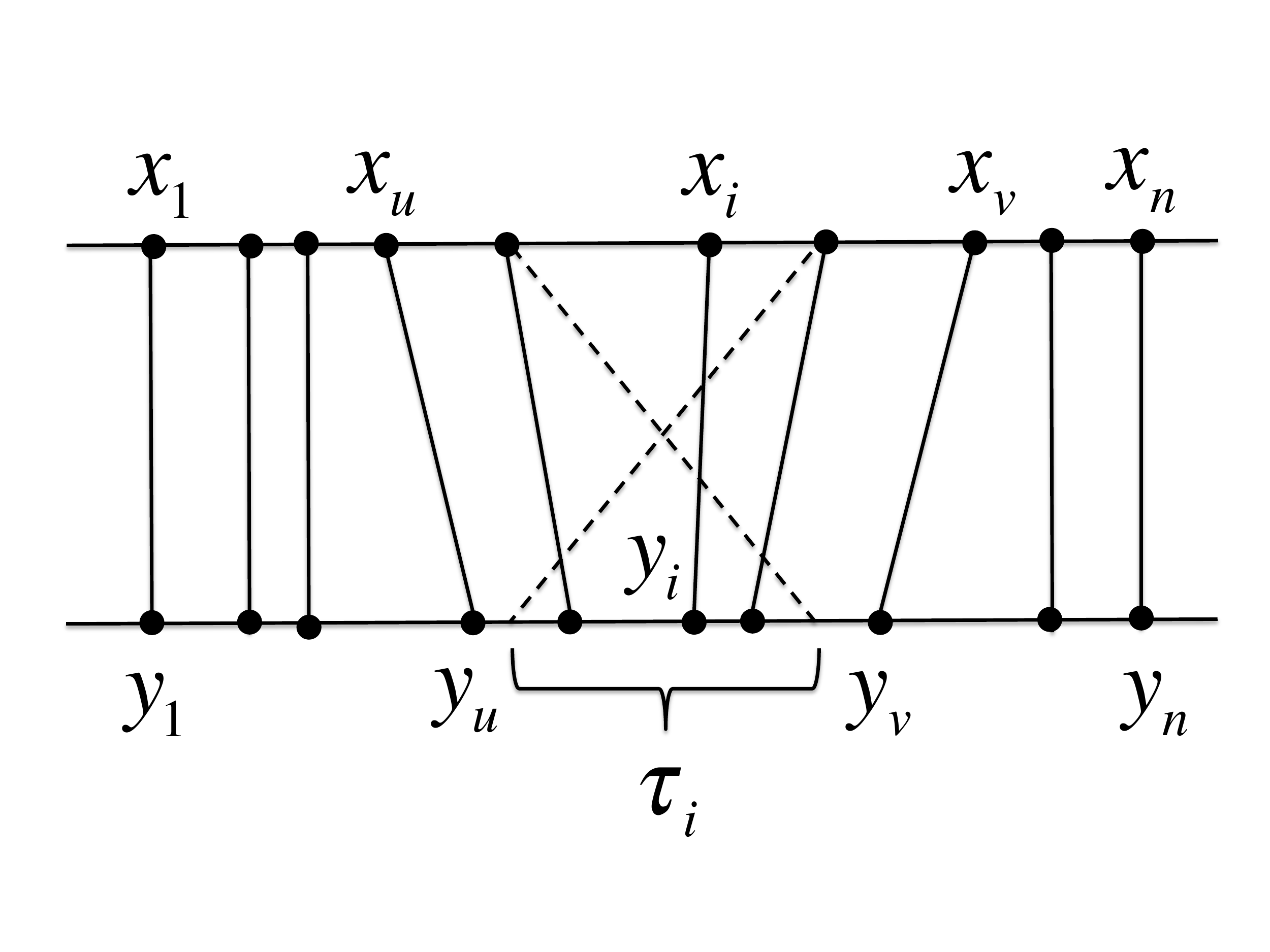}
\end{center}
\vspace{-0.8cm}
\caption{\small  The interval $\tau_i$ extends from a distance $\rho(x_{i+1}-x_u)$
to the right of $x_u$ to a distance $\rho(x_v-x_{i-1})$ to the left of $x_v$: 
it thus {\em twists} $[x_{i-1},x_{i+1}]$ into the allowed interval for $y_i$.
\label{fig1}}
\end{figure}

\smallskip

Observe that conditions (i,ii) are always feasible:  
for example, we can choose $y_i$ to be the leftmost point in $\tau_i$;
of course, there is no need to do so and the expressive power of twist systems 
comes from the freedom they offer.
Like their averaging counterparts,
such systems are highly nondeterministic: at each step $t$, both 
the choice of $u,v$ and the motion of the points are 
entirely arbitrary within the constraints (i,ii). 
Writing $\ell_t= (x_v-x_u)^s$, we define 
the $s$-energy of the twist system as ${\mathcal E} (s) =  \sum_{t> 0} \ell_t$.
The next result justifies the introduction of twist systems.

\begin{theorem}\label{FromAvetoTwist}
$\!\!\! .\,\,$
Any averaging system can be viewed as a twist system with
the same parameter $\rho$ and the same $s$-energy.
\end{theorem}

\proof
Referring to our previous notation,  
recall that $\mu_1,\ldots, \mu_k$ denote the lengths of the intervals $I_j$
formed by the union of the edge embeddings of $g_t$.
We subdivide the time interval from $t$ to $t+1$ into $k$ time windows
and, for $j=1,\ldots, k$, we process the motion within $I_j$ during the $j$-th window
while keeping the other vertices fixed. All windows are treated similarly,
so it suffices to explain the case $k=1$. 
Let $x_i$ (resp. $x_i'$) be the position of vertex $i$ at time $t$ (resp. $t+1$)
and let $y_1\leq\cdots\leq y_n$ be the sequence of $x_i'$ sorted in nondecreasing 
order.\footnote{We break ties by using the index $i$. Note that the $y_i$'s are sorted,
so they are not the same as those used in the definition of averaging systems given above.}
Let $x_u,\ldots, x_v$ denote the positions within $I_1$;
we may assume that $u<v$.
The other vertices are kept fixed, so we have $y_i=x_i$ for $i<u$ or $i>v$.
To show that the transition from $x_i$ to $y_i$ meets the conditions of
a twist system, we need to prove that 
$y_i\in \tau_i$ for any $i$ between $u$ and $v$.
By the symmetry of~(\ref{twistDef}), it suffices to show that, for $u\leq i\leq v$,

\begin{equation}\label{tauCond}
y_i\leq x_v- \rho (x_v- x_{\max\{i-1,u\}}).
\end{equation}
Assume that $u<i\leq v$
and let $\bar x_j$ be shorthand for $ \rho x_j + (1-\rho)x_v$.
The entire interval $I_1$ is covered by edges of $g_t$,
so there must be at least one edge $(a,b)$ that covers $[x_{i-1},x_i]$, ie,
$b < i \leq a$.
By~(\ref{iter}),
$x_a'\leq \rho x_{l(a)}+ (1-\rho) x_{r(a)}$, with $l(a)\leq b <i$ and $r(a)\leq v$;
hence $x_a'\leq \bar x_{i-1}$.
It also follows from~(\ref{iter}) and the presence of self-loops 
that $x_j'\leq \rho x_{l(j)}+(1-\rho) x_{r(j)} \leq \bar x_{i-1}$
for any $j$ ($u\leq j<i$); also $x_j'=x_j\leq  \bar x_{i-1}$ for $j<u$.
Putting it all together, this 
proves the existence of at least $i$ indices $l \leq v$ such that
$x_l'\leq \bar x_{i-1}$. It follows that $y_i\leq \bar x_{i-1}$; hence~(\ref{tauCond}) for
$u<i\leq v$.  To complete the proof of~(\ref{tauCond}),
we note that the case $i=u$ follows from $y_u\leq y_{u+1}$. The case $k>1$ is handled 
by repeating the previous analysis for each interval $I_j$. The $s$-energy
contributed by one step of the averaging system matches the 
energetic contribution of the $k$ substeps of the twist system.
\hfill $\Box$
\proofend

\section{Bounding the $s$-Energy}\label{BoundsEnergy}

The proof of Theorem~\ref{s-energyBound}
is unusual in the context of dynamics because it is 
{\em algorithmic}: it consists of a set of trading rules 
that allows money to be injected into the system and exchanged among the vertices
to meet their needs. As the transactions take place, money is spent to pay for the
$s$-energy expended along the way. If all of the energy can be
accounted for in this manner, then the amount of money injected in
the system is an upper bound on $\mathcal{E}(s)$.  
In our earlier work~\cite{chazelle-total}, we were able to
pursue this approach only for the case $s=1$.  We show here how to 
extend it to all $s\in (0,1]$. 
The idea was to supply each vertex with its own credit account
and then let them trade credits to pay for the $s$-energy incrementally.
This strategy does not work here because
of its inability to cope with all the scales present in the system.\footnote{We illustrate
the difficulty with a simple example. Set $n=3$ 
and assign $x_i^sA^i$ credits to the account for vertex $i=1,2,3$.
Initialize the system with $x_1=0$, $x_2=1-\eps$, and $x_3=1$;
set $\rho= 1/2$, with $g_1$ consisting of the single edge $(2,3)$.
Assume now that $y_1=0$ and $y_2=  y_3= 1-\eps/2$. The account for vertex 3, the only one
to release money, gives out only $(1- (1-\eps/2)^s)A^3\approx \frac{1}{2}s \eps A^3 $ credits.
If $s<1$ and $\eps>0$ is very small, this is not enough 
to cover the $s$-energy of $\eps^s$ needed for the first step.
The problem is that the credit accounts do not operate at all scales.}
The remedy is to supply each {\em pair} of vertices with their own account.
Only then are we able to accommodate all scales at once.
By appealing to Theorem~\ref{FromAvetoTwist}, we may substitute
twist systems for averaging systems.
We focus the analysis on the transition at time $t$
from $x_1\leq\cdots\leq x_n$ to $y_1\leq\cdots\leq y_n$.
Our only assumption is that, for some $u, v$ ($1\leq u<v\leq n$), we have 
$y_i\in \tau_i$ for any $u\leq i\leq v$, and $y_i=x_i$ otherwise.

For each pair $(i,j)$ such that $1\leq i<j\leq n$, we maintain an account $B_{i,j}$ consisting
of $(x_j-x_i)^sA^{j-i}$ credits,
where $A\!: = 2/\rho s$ and one credit is used to pay for a single unit of  
$s$-energy. (Amounts paid need not be integers.)
We show that updating each $B_{i,j}$ at time $t$ to $B_{i,j}'$ at time $t+1$
leaves us with enough unused money to pay for the $s$-energy $(x_v-x_u)^s$  released
at that step.\footnote{We refer to $B_{i,j}$ as both the account for $(i,j)$ and its value.}
No new money is needed past the initial injection at time 1, so
the $s$-energy is at most the sum of all the $B_{i,j}$'s at the beginning:
${\mathcal E}(s)\leq 
\sum_{i<j}A^{j-i}< \bigl(\frac{A}{A-1}\bigr)^2 A^{n-1}
< 2(2/\rho s)^{n-1}$, for $n>2$. For $n=2$, ${\mathcal E}(s)\leq A$, hence
Theorem~\ref{s-energyBound}.
We begin with a few words of intuition:

\begin{itemize}
\item
We update $B_{i,j}$ to $B'_{i,j}$
by considering the pairs $(i,j)$ in descending order of $j-i$, starting with $(1,n)$.
In general, the update for $(i,j)$ will rely on money
released by the pairs $(i-1,j)$ and $(i,j+1)$, whose accounts
will have already been updated.
In turn, the pair $(i,j)$ will then be 
expected to provide money to both $(i,j-1)$ and $(i+1,j)$:
the donation will be made in two equal amounts.

\item 
How much money should $(i,j)$ receive from its donors.
For the sake of this informal discussion, let us focus on the case $u\leq i<j\leq v$.
The account $B_{i,j}$ should receive enough to grow to
$(x_v-x_u)^s A^{j-i}$. This typically exceeds its balance of 
$(x_j-x_i)^s A^{j-i}$ at time $t$, so an infusion of money is required.
Of course, the amount actually needed for $B'_{i,j}$ 
is only $(y_j-y_i)^s A^{j-i}$, so this in turn frees
$\bigl( (x_v-x_u)^s - (y_j-y_i)^s \bigr) A^{j-i}\geq 0$, which can be then
passed on to $(i,j-1)$ and $(i+1, j)$.

\item
We pay for the energetic contribution at time $t$
by spending the leftover money from the update for $(u,u+1)$, which
we show to be at least $(x_v-x_u)^s$, as required.
\end{itemize}

\vspace{-0.1cm}
\begin{figure}[htb]
\begin{center}
\hspace{0cm}
\includegraphics[width=8cm]{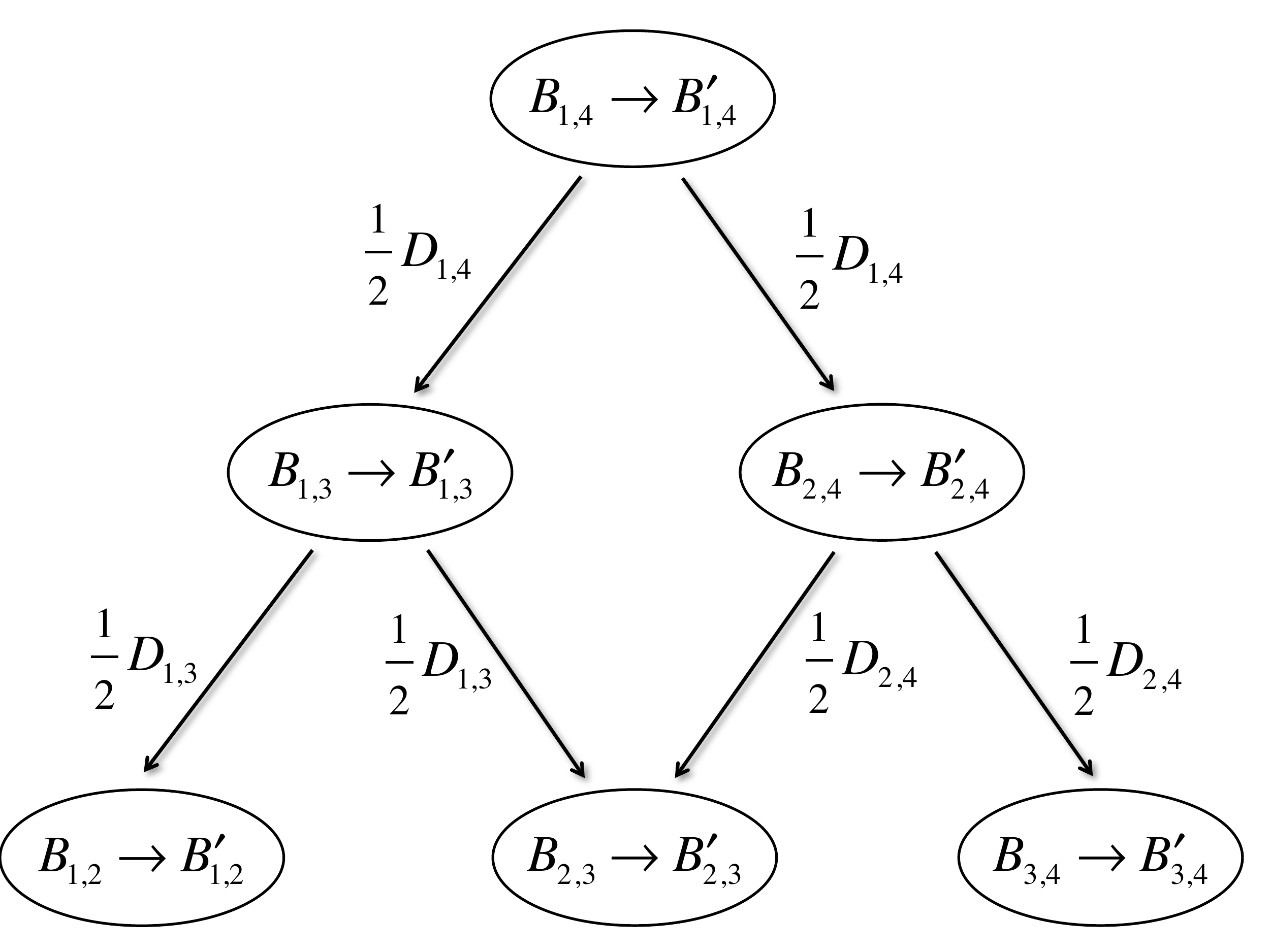}
\end{center}
\vspace{-0.4cm}
\caption{\small  Updating $B_{1,4}$ to its new value of $B'_{1,4}$
releases $D_{1,4}$ credits, which are passed on evenly to the pairs $(1,3)$ and 
$(2,4)$. With this scheme in place, updating $B_{2,3}$ to $B'_{2,3}$ can make use
of $C_{2,3}= \frac{1}{2}(D_{1,3}+ D_{2,4})$ credits.
\label{fig2}}
\end{figure}

\bigskip
\noindent
{\em Proof of Theorem~\ref{s-energyBound}}.
We update $B_{i,j}$ by using $C_{i,j}$ credits supplied
by the accounts $B_{i-1,j}$ and $B_{i,j+1}$.
We show how this produces a leftover $D_{i,j}$, which can then be donated
to $(i+1,j)$ and $(i,j-1)$ in equal amounts.
Here are the details: for all $1\leq i< j\leq n$ in descending order of
$j-i=n-1,\ldots, 1$, apply the following assignments (Fig.2):

\begin{equation}\label{clearing}
\begin{cases}
\,
C_{i,j} \leftarrow \, \frac{1}{2}(D_{i-1,j}+ D_{i,j+1}) \\
\,
D_{i,j} \leftarrow \, B_{i,j} +  C_{i,j} - B'_{i,j}  ,
\end{cases}
\end{equation}
where $B_{i,j}= (x_j-x_i)^sA^{j-i}$,
$B'_{i,j} =  (y_j- y_i)^sA^{j-i}$,
and $D_{i,j}=0$ if $i<1$ or $j>n$. 
The assignments denote transfers of money. This explains the
factor of $1/2$, which keeps the money pool conserved:
for example, one half of $D_{i,j}$ goes to
$(i,j-1)$ and the other half to $(i+1,j)$.
The soundness of the trading scheme rests entirely
on the claimed nonnegativity of all the donations $D_{i,j}$. 
For any $i\in \{1,\ldots, n\}$,
define $u(i)=u$ and $v(i)=v$ if $u\leq i\leq v$; and set $u(i)=v(i)=i$ otherwise.
We prove by induction on $j-i>0$ that, for $1\leq i<j \leq n$,
\begin{empheq}[left=\empheqlbrace]{align}
&\,
B_{i,j}+C_{i,j}\geq (x_{v(j)} - x_{u(i)})^s A^{j-i} \label{BClb}\\
&\,
D_{i,j}\geq 0. \label{Dlb} 
\end{empheq}

\noindent
The next inequality, which follows from
$d z^s/dz \geq s$ for $s,z\in (0,1]$,
will prove useful in establishing~(\ref{BClb}, \ref{Dlb}):
\begin{equation}\label{ineqs}
1- (1-x)^s \geq sx  \hspace{0.3cm}\text{for any $s, x\in [0,1]$.}
\end{equation}

\bigskip
\noindent
$\bullet$ {\em Case $u\leq  i<j \leq v$}.
By affine invariance, we can always assume that $x_u= x_v-1=0$.
We begin with the case $u<i<j\leq v$ 
and observe that $v(j)=v$ and $u(i-1)=u(i)= u$.
Because $y_{i-1}\in \tau_{i-1}$, we have $y_{i-1}\geq \rho x_i$. 
Using~(\ref{ineqs}), we find that
\begin{equation}\label{Di-1jLBnd}
\begin{split}
D_{i-1,j}
&= B_{i-1,j}+C_{i-1,j} - B'_{i-1,j} 
\geq
\bigl(   (x_{v(j)} - x_{u(i-1)})^s - (y_j - y_{i-1})^s\bigr)A^{j+1-i} \\
&\geq 
\bigl(1- (1 - \rho x_i)^s\bigr) A^{j+1-i} 
\geq \rho s x_i A^{j+1-i}.
\end{split}
\end{equation} 
If $i=u$, we have $x_i=0$, hence (\ref{Di-1jLBnd}) merely expresses
nonnegativity, which holds inductively.
We conclude that~(\ref{Di-1jLBnd}) obtains for any $u\leq i<j\leq v$. Likewise, 
by symmetry,
$D_{i,j+1} \geq \rho s(1-x_j) A^{j+1-i}$.
It follows from~(\ref{clearing}) that
$C_{i,j}\geq \frac{1}{2}\rho s\bigl(1- (x_j-x_i)\bigr) A^{j+1-i}
\geq  \bigl(1- (x_j-x_i)^s\bigr) A^{j-i}$;
therefore,
$$B_{i,j}+C_{i,j}\geq  
        (x_j-x_i)^s  A^{j-i} 
   +   \bigl(1- (x_j-x_i)^s\bigr) A^{j-i}
   = A^{j-i} = (x_{v(j)} - x_{u(i)})^s A^{j-i},$$
which establishes~(\ref{BClb}).
Since $x_{u(i)}= x_u \leq y_i \leq y_j\leq x_v= x_{v(j)}$,
this also proves that 
$$D_{i,j} = B_{i,j}+C_{i,j} - B'_{i,j}
\geq 
A^{j-i} - (y_j - y_i)^s A^{j-i}  \geq 0; $$
hence~(\ref{Dlb}).

\bigskip
\noindent
$\bullet$ {\em Case $i<u\leq j\leq v$}. This time, we set $x_i=0$ and $x_v=1$
and note that $u(i)=i$ and $v(j)=v$. 
We begin with the case $j<v$,
which implies that $v(j+1)=v$.
Using~(\ref{clearing}, \ref{BClb}), $y_i =x_i$, 
$y_{j+1}\in \tau_{j+1}$, 
and~(\ref{ineqs}) in this order, we find that
\begin{equation}\label{dij2}
\begin{split}
D_{i,j+1}&= B_{i,j+1}+C_{i,j+1} - B'_{i,j+1} \geq
\bigl(   (x_{v(j+1)} - x_{u(i)})^s - (y_{j+1} - x_i)^s\bigr)A^{j+1-i} \\
&\geq \bigl( 1- (1-\rho (1-x_j))^s\bigr) A^{j+1-i}
\geq \rho s (1-x_j) A^{j+1-i}
\geq  (1-x_j) A^{j-i}.
\end{split}
\end{equation}
Again, by induction, $D_{i-1,j}\geq 0$; therefore, by~(\ref{clearing}),
$$B_{i,j}+C_{i,j}\geq  B_{i,j}+  \hbox{$\frac{1}{2}$} D_{i,j+1}
\geq  \bigl(  x_j^s  +   (1-x_j) \bigr)A^{j-i} 
\geq A^{j-i} =  (x_{v(j)} - x_{u(i)})^s A^{j-i};$$
hence~(\ref{BClb}).
For the case $j=v$, again note that the lower bounds 
on $D_{i-1,j}$ and $D_{i,j+1}$ we just used still hold,
and thus so does~(\ref{BClb}) for all $i<u\leq j\leq v$.
Finally, $y_j\leq x_v$; hence
$x_{u(i)}= x_i= y_i \leq y_j\leq x_v= x_{v(j)}$,
and~(\ref{Dlb}) follows from~(\ref{clearing}, \ref{BClb}).

The case $u\leq i\leq v <j$ is the mirror image of the last one
while the remaining three cases are trivial and require no account updates.
We pay for the $s$-energy contribution at time $t$
by tapping into $D_{u,u+1}$, which is unused.
For this to work, it suffices to 
show that $D_{u,u+1}\geq (x_v-x_u)^s$. We have
$y_i\in \tau_i$ ($i=u, u+1$); hence
\begin{equation*}
\begin{split}
y_{u+1} - y_u 
&\leq x_v -\rho (x_v-x_u) - ( x_u+ \rho (x_{u+1}-x_u) ) \\
&\leq \rho (x_u- x_{u+1})  + (1-\rho)(x_v - x_u)
        \leq (1-\rho)(x_v - x_u).
\end{split}
\end{equation*}
Thus, it follows from~(\ref{clearing}, \ref{BClb}, \ref{ineqs}),
together with $u(u)=u$ and $v(u+1)=v$, that 
\begin{equation*}
\begin{split}
D_{u,u+1}
&=  B_{u,u+1}+C_{u,u+1}-B'_{u,u+1}
\geq 
A ( x_{v(u+1)}- x_{u(u)} )^s 
-  A ( y_{u+1} - y_{u} )^s \\
&\geq \bigl( 1 - (1-\rho)^s \bigr) A(x_v-x_u)^s 
\geq \rho s A (x_v-x_u)^s\geq (x_v-x_u)^s.
\end{split}
\end{equation*}
This completes the proof of Theorem~\ref{s-energyBound}
for twist systems. 
By Theorem~\ref{FromAvetoTwist}, this also implies the same upper bound
for averaging systems.
\hfill $\Box$
\proofend

\section{Applications}

A number of known convergence rates for various averaging systems
can be sharpened by appealing to Theorems~\ref{s-energyBound} and~\ref{CommCount}.
We give a few examples below.

\subsection{Asymmetric averaging systems}

Symmetric averaging systems have been widely used to model
backward products of the form $(A_t\cdots A_1 x)_{t>0}$,
where each $A_k$ is a type-symmetric stochastic matrix with positive diagonal
and nonzero entries at least $\rho>0$~\cite{caoSM, FagnaniF, HendrickxB, jadbabaieLM03,
lorenz05, Moreau2005}.\footnote{A matrix
$A$ is type-symmetrix if
$A_{ij}$ and $A_{ji}$ are both positive or both 0 for all $i,j$.}
In other words, $A_k$ is the matrix of a lazy random walk 
in an undirected graph $g_k$ with a lower bound of $\rho$ on the
nonzero probabilities. 
A close examination of the proof of Theorem~\ref{FromAvetoTwist}
shows that the graphs $g_t$ may be directed as long as
the vertices still have self-loops,
and, for each $i=u+1,\ldots, v$, 
there exist edges ``hovering" over $i$ from both sides, ie,
$(a,b)$ and $(b',a')$, with $a, a' < i \leq b,b'$. We note that this property holds if
each directed graph $g_t$ is cut-balanced.\footnote{A directed
graph is cut-balanced if its weakly connected components
are strongly connected.} This gives us a strict generalization of 
Theorems~\ref{s-energyBound} and~\ref{CommCount} for
asymmetric averaging systems whose sequences of digraphs
are cut-balanced. This goes beyond the convergence of these systems,
which was established in~\cite{HendrickxT13}.

\subsection{Opinion dynamics}

There has been considerable attention given to
consensus formation in social dynamics~\cite{FagnaniF, fortunato05, hegselmanK}.
Given a set of agents in high-dimensional space, where
coordinates model opinions, one imagines
that at each step a subset
of them come into contact and, through a process of deliberation,
adjust their opinions toward agreement.
Will such a process converge to consensus, polarization,
a mixture of both, or not at all?
Mathematically, the agents are represented by
their position in $d$-dimensional space:
$x_1,\ldots, x_n$ in $[0,1]^d$.
We fix $0<\alpha\leq 1$ and iterate on the following process forever:
(1) choose an arbitrary nonempty subset of the agents and move them
anywhere inside the box $(1-\alpha)B + \alpha c$, where $B$
is the smallest (axis-parallel) box enclosing the chosen agents and $c$ is 
the center of $B$; (2) repeat.
Intuitively, one ``squeezes" the subset of agents together a little.

\begin{theorem}\label{OpinionDyn}
$\!\!\! .\,\,$
For any positive $\eps \leq 2^{-dn}$, at all but
$O\bigl(\frac{1}{d\alpha n}\log \frac{1}{\eps}\bigr)^{n-1}$ time steps,
the smallest box enclosing the chosen agents has volume less than $\eps$.
\end{theorem}

\proof
We set up a symmetric averaging system as follows:
$g_t$ consists of $n$ self-loops, together with 
the complete graph joining the agents of the chosen subset;
along each axis, the dynamics obeys~(\ref{iter}) with parameter $\rho= \alpha/2$.
Let $\ell_t(j)$ be the length of the graph's projection onto 
the $j$-th axis. By Theorem~\ref{s-energyBound},
we know that, for any $0<r\leq 1$,
$\sum_{t>0} \ell_t(j)^r \leq  (6/\alpha r)^{n-1}$.
Let $V_t$ be the volume of the smallest box enclosing the agents picked at time $t$.
By the generalized H\"older's inequality, for $0<s\leq 1/d$,
$$
\sum_{t>0} V_t^s =
\sum_{t>0} \prod_{j=1}^d \ell_t(j)^s \leq  
 \prod_{j=1}^d  \, \Bigl( \sum_{t>0}  \ell_t(j)^{ds} \Bigr)^{1/d}
\leq (6/d\alpha s)^{n-1}.
$$
Set $s= n/\log \frac{1}{\eps}$ and use Markov's inequality to complete the proof.
\hfill $\Box$
\proofend

\subsection{Flocking}

Many models of bird flocking have been developed 
over the years and used to great effect in CGI for film and animation.
Their mathematical analysis has lagged behind, however.
In a simple, popular model tracing its roots back to Cucker \& Smale, 
Vicsek, and ultimately Reynolds, a group of $n$ birds is
represented by two $n$-by-$3$ matrices $\mathbf{x}(t)$ 
and $\mathbf{v}(t)$, where the $i$-th rows
encode the location and velocity in $\mathbb{R}^3$ of the $i$-th 
bird, respectively~\cite{CuckerSmale1, jadbabaieLM03, vicsekCBCS95}.
The dynamics obeys the relations

\begin{equation*}\label{modelD}
\begin{cases}
\,\, \mathbf{x}(t)= \mathbf{x}(t-1)+ v(t) \\
\,\, \mathbf{v}(t+1)= P(t)\, \mathbf{x}(t),
\end{cases}
\end{equation*}
where $P(t)$ is an $n$-by-$n$ stochastic matrix whose entry
$(i,j)$ is positive if and only if birds~$i$ and~$j$ are within a fixed
distance $R$ of each other. All entries are rationals
over $O(\log n)$ bits. A tight bound on the convergence of
the dynamical system was established in~\cite{chazFlockPaperUB, chazFlockPaperLB}: 
it was shown that steady state is always reached within a number of steps
equal to a tower-of-twos of height proportional to $\log n$;
even more amazing, this bound is optimal.
The lead-up to steady-state consists of two phases:
fragmentation and aggregation. The latter can feature only
the merging of flocks while the (much shorter) fragmentation phase 
can witness the repeated formation and breakup of flocks.
Technically, a flock is defined as the birds in a given connected
component of the network joining any two birds
within distance $R$.  It has been shown that the total number
of network switches (ie, the number of steps where
the communication network changes)
is $n^{O(n^2)}$.   We improve this bound to
$n^{O(n)}$ by using the $s$-energy. It was demonstrated
in~\cite{chazFlockPaperUB} (page 21:7) that the
number of network switches is bounded by 
the communication count $C_\eps$, for $\eps\geq n^{-bn^2}$, $\rho\geq n^{-c}$
and constant $b, c>0$. Our claim follows from Theorem~\ref{CommCount}.
\hfill $\Box$
\proofend

\subsection{Self-synchronizing oscillators}

The self-organized synchronization of coupled oscillators 
is a well-known phenomenon in physics and biology: it is
observed in circadian neurons,
firing fireflies, yeast cell suspensions, cardiac pacemaker cells, 
power plant grids, and even musical composition (eg,
Ligeti's {\em po\`eme symphonique}).
In the discrete Kuramoto model studied 
in~\cite{martinez09,Moreau2005,scardoviSS},
all oscillators share the same natural frequency
and the phase of the $i$-th one obeys the recurrence:
\begin{equation*}
\theta_i(t+1)= \theta_i(t) 
+ \frac{K\Delta T}{|n_i(t)|}
\sum_{j\in n_i(t)}\sin\bigl(\theta_j(t)-\theta_i(t)\bigr),
\end{equation*}
where $n_i(t)$ is the set of vertices adjacent to $i$ in $g_t$ (which includes $i$). 
Following~\cite{Moreau2005}, we assume
that all $n$ phases start in the same open half-circle, which we can express
as $\alpha-\pi/2\leq \theta_i(0)\leq \pi/2$, for
some arbitrarily small positive constant $\alpha$.
We find that
$\sin(\theta_j(0)-\theta_i(0)) = a_{ij}\bigl(\theta_j(0)-\theta_i(0)\bigr)$,
where $c \alpha\leq a_{ij}\leq 1$, for constant $c>0$. 
This condition holds for all $t$ since averaging keeps
the phases in the same open half-circle.  The dynamics is
that of a symmetric
averaging system provided that we pick $\rho$ small enough so that 
$b\rho n/\alpha \leq K\Delta T \leq 1$, for a suitable constant $b>0$.
By Theorem~\ref{CommCount}, for any $\eps\leq 2^{-n}$,
the number of steps where two oscillators are joined by
an edge while their phases are off by $\eps$ or more is
$O\bigl(\frac{1}{\alpha K \Delta T}\log \frac{1}{\eps}\bigr)^{n-1}$.

\section{The Lower Bound Proofs}\label{lbProofs}

\bigskip
\noindent
{\bf A.}\ 
We prove that the bound $O(1/\rho s)^{n-1}$ from Theorem~\ref{s-energyBound} is optimal 
for $s= O\bigl( 1/\log \frac{1}{\rho} \bigr)$ and $\rho \leq 1/3$.
A lower bound construction from~\cite{chazelle-total}
(page~1703) describes a system whose $n$-agent $s$-energy satisfies the recurrence 
$\mathcal{E}_n\geq \rho^s \mathcal{E}_{n-1}+ (1-2\rho)^s\mathcal{E}_n +1$
for $n>1$; hence, for positive constant~$b$,  
$\mathcal{E}_2\geq b/\rho s$ and
$\mathcal{E}_n\geq (b \rho^{s-1}/s) \mathcal{E}_{n-1}$ for $n>2$.
This shows that 
$\mathcal{E}_n\geq (b/\rho s)^{n-1}\rho^{s(n-2)}= 
\Omega(1/\rho s)^{n-1}$,  for $s= O\bigl(1/\log \frac{1}{\rho} \bigr)$, as claimed.
\hfill $\Box$
\proofend

\bigskip
\noindent
{\bf B.}\ 
We prove that
$\mathcal{C}_\eps=  \Omega \Bigl( \frac{1}{\rho n} \log \frac{1}{\eps} \Bigr)^{n-1}$,
for any positive $\eps\leq \rho^{2n}$ and $\rho\leq 1/3$.
Note that $\rho$ must be bounded away from $1/2$ (we choose $1/3$ for convenience):
indeed, in the case of two vertices at distance~1 joined by an edge,
we have the trivial bound $C_\eps=1$ for $\rho=1/2$.
The proof revisits an earlier construction~\cite{chazelle-total}
and modify its analysis to fit our purposes.
If $n>1$, the $n$ vertices of $g_1$ are positioned at 0, except for $x_n=1$.
Besides the self-loops, the graph $g_1$ has the single edge 
$(n-1,n)$. At time 2, the vertices are all at $0$ except for
$x_{n-1}=\rho$ and $x_n= 1-\rho$. 
The first $n-1$ vertices form a system that stays in place if $n=2$ and, otherwise,
proceeds recursively within $[0,\rho]$ until it converges to the fixed point $\rho/(n-1)$:
this value is derived from the fact that each step keeps
the mass center invariant.
After convergence\footnote{We can use a limiting argument
to break out of the infinite loop.}
of the vertices labeled $1$ through $n-1$,
the $n$-vertex system repeats the previous construction within $[\rho/(n-1), 1-\rho]$.
Let $C(n,\eps)$ denote the communication count for $n$ agents:
we have $C(n,\eps)=0$ if $n=1$ or $\eps>1$; else 
\begin{equation}\label{recurrenceGeq1}
C(n, \eps) \geq 1 + C\, \biggl (\, n-1, \frac{\eps}{\rho} \, \biggr) + 
C\, \biggl( \, n,  \frac{\eps}{1- \rho n/(n-1)} \, \biggr)  \, .
\end{equation}
By expanding the recurrence and using monotonicity,
\begin{equation}\label{recurrenceGeq-k}
C(n, \eps) \geq k+  k\,  C \, \biggl(\, n-1,  \frac{\eps}{\rho(1- 2\rho)^{k-1}} \, \biggr)\,\, ,
\hspace{.3cm} \text{for}\,\,\, 
k= \biggl \lceil \frac{(\log \eps)/n-\log\rho}{2\log(1-2\rho)} \biggr\rceil.
\end{equation}
Assume now that $\eps \leq \rho^{2n}$.
From our choice of $k$, we easily verify that
\begin{equation}\label{kBnd}
\rho(1- 2\rho)^{k-1} \geq \eps^{1/n} \, .
\end{equation}
The recurrence~(\ref{recurrenceGeq-k}) requires
that $\eps/\bigl( \rho(1- 2\rho)^{k-1} \bigr) <1$, which follows from~(\ref{kBnd}).
Since $\eps^{1/2n} \leq \rho \leq 1/3$, we have $k\geq \frac{b}{\rho n}\log \frac{1}{\eps}$,
for constant $b>0$.
It follows that $C(2,\eps)= \Omega(\frac{1}{\rho}\log \frac{1}{\eps})$
and, for $n>2$, by~(\ref{kBnd}),
\begin{equation*}
C(n, \eps) \geq \biggl(\, \frac{b}{\rho n}\log \frac{1}{\eps} \, \biggr) \,\,
C\Bigl( n-1,  \eps^{1-1/n} \Bigr).
\end{equation*}
We verify that the condition $\eps\leq \rho^{2n}$ holds recursively:
$\eps^{1-1/n} \leq  \rho^{2(n-1)}$.
By induction, it follows that
$C(n, \eps) \geq \Omega\Bigl( \frac{1}{\rho n}\log \frac{1}{\eps} \Bigr)^{n-1}$,
as desired.
\hfill $\Box$
\proofend

\bigskip



\end{document}